%% file: conf_sum2004.tex
\documentclass[11pt]{article}
\usepackage{epsfig}
\usepackage{graphicx}
\usepackage{subfigure}
\usepackage{dcolumn}
\usepackage{bm}

\newcommand{\BABARPubYear}    {04}

\newcommand{\BABARConfNumber} {16}
\newcommand{\SLACPubNumber} {10675}
\newcommand{\LANLNumber} {0000}

\long\def\inst#1{\par\nobreak\kern 4pt\nobreak
    {\it #1}\par\vskip 10pt plus 3pt minus 3pt}
\usepackage{relsize}
\usepackage{xspace}
\def\babar{\mbox{\slshape B\kern-0.1em{\smaller A}\kern-0.1em
    B\kern-0.1em{\smaller A\kern-0.2em R}}}
\def\pep2{PEP-II}
\def\CP{\ensuremath{C\!P}\xspace}

\def\pipm  {\ensuremath{\pi^\pm}\xspace}

\def\Kbar  {\kern 0.2em\overline{\kern -0.2em K}{}\xspace}

\def\Kstarzb {\ensuremath{\Kbar^{*0}}\xspace}

\def\Dp      {\ensuremath{D^+}\xspace}

\def\Ds      {\ensuremath{D^+_s}\xspace}
\def\mphi       {\mbox{$\phi$}\xspace}
\def\Dptokkpi   {\ensuremath{\Dp \to K^{-}K^{+}\pi^{+}}\xspace}

\def\Dptokpipi  {\ensuremath{\Dp \to K^{-}\pi^{+}\pi^{+}}\xspace}

\def\Dptophipi  {\ensuremath{\Dp \to \mphi\pi^{+}}\xspace}
\def\Dptokstark {\ensuremath{\Dp \to \Kstarzb K^{+}}\xspace}

\def\Dsptokkpi  {\ensuremath{\Ds \to K^{-}K^{+}\pi^{+}}\xspace}

\newcommand{\gevc}{\ensuremath{{\mathrm{\,Ge\kern -0.1em V\!/}c}}\xspace}
\newcommand{\mevc}{\ensuremath{{\mathrm{\,Me\kern -0.1em V\!/}c}}\xspace}
\newcommand{\gevcc}{\ensuremath{{\mathrm{\,Ge\kern -0.1em V\!/}c^2}}\xspace}
\newcommand{\mevcc}{\ensuremath{{\mathrm{\,Me\kern -0.1em V\!/}c^2}}\xspace}
\def\invfb   {\ensuremath{\mbox{\,fb}^{-1}}\xspace}

\begin{document}
{\pagestyle{empty}

\begin{flushright}
\babar-CONF-\BABARPubYear/\BABARConfNumber \\
SLAC-PUB-\SLACPubNumber \\
hep-ex/\LANLNumber \\
August 2004 \\
\end{flushright}

\par\vskip 5cm

\begin{center}
\Large \bf \boldmath \Dptokkpi Meson Decays: A Search for $C\!P$ Violation
and a Measurement of the Branching Ratio
\end{center}
\bigskip

\begin{center}
\large The \babar\ Collaboration\\
\mbox{ }\\
\today
\end{center}
\bigskip \bigskip

\begin{center}
\large \bf Abstract
\end{center}
We present a preliminary measurement of the \CP asymmetry in singly Cabibbo-suppressed \Dptokkpi
decays and in the resonant decays \Dptophipi and \Dptokstark . We use a data sample of $79.9$ \invfb recorded by the \babar\ detector. 
The Cabibbo-favored \Dsptokkpi branching fraction is used as normalization  
in the measurements to reduce systematic uncertainties. 
Preliminary results of the \CP asymmetries obtained are
$A_{CP}(K^{+}K^{-}\pipm) = (1.4\pm1.0(\textrm{stat.})\pm1.1(\textrm{syst.}))\times 10^{-2}$,
$A_{CP}(\mphi\pipm) = (0.2\pm1.5(\textrm{stat.})\pm0.8(\textrm{syst.}))\times 10^{-2}$, and
$A_{CP}(K^{\pm}\Kstarzb) = (0.9\pm1.7(\textrm{stat.})\pm0.8(\textrm{syst.}))\times 10^{-2}$.
A preliminary determination of the branching ratio is $\frac{\Gamma (\Dptokkpi )}{\Gamma(\Dptokpipi)}=(10.7\pm0.1(\textrm{stat.})\pm0.2(\textrm{syst.}))\times 10^{-2}$.

\vfill
\begin{center}

Submitted to the 32$^{\rm nd}$ International Conference on High-Energy Physics, ICHEP 04,\\
16 August---22 August 2004, Beijing, China

\end{center}

\vspace{1.0cm}
\begin{center}
{\em Stanford Linear Accelerator Center, Stanford University, 
Stanford, CA 94309} \\ \vspace{0.1cm}\hrule\vspace{0.1cm}
Work supported in part by Department of Energy contract DE-AC03-76SF00515.
\end{center}

\newpage
} 

%
\input pubboard/authors_sum2004.tex

\section{INTRODUCTION}
\label{sec:Introduction}
\begin{figure}
  \centering
  \subfigure[]{\includegraphics[width=.49\columnwidth]{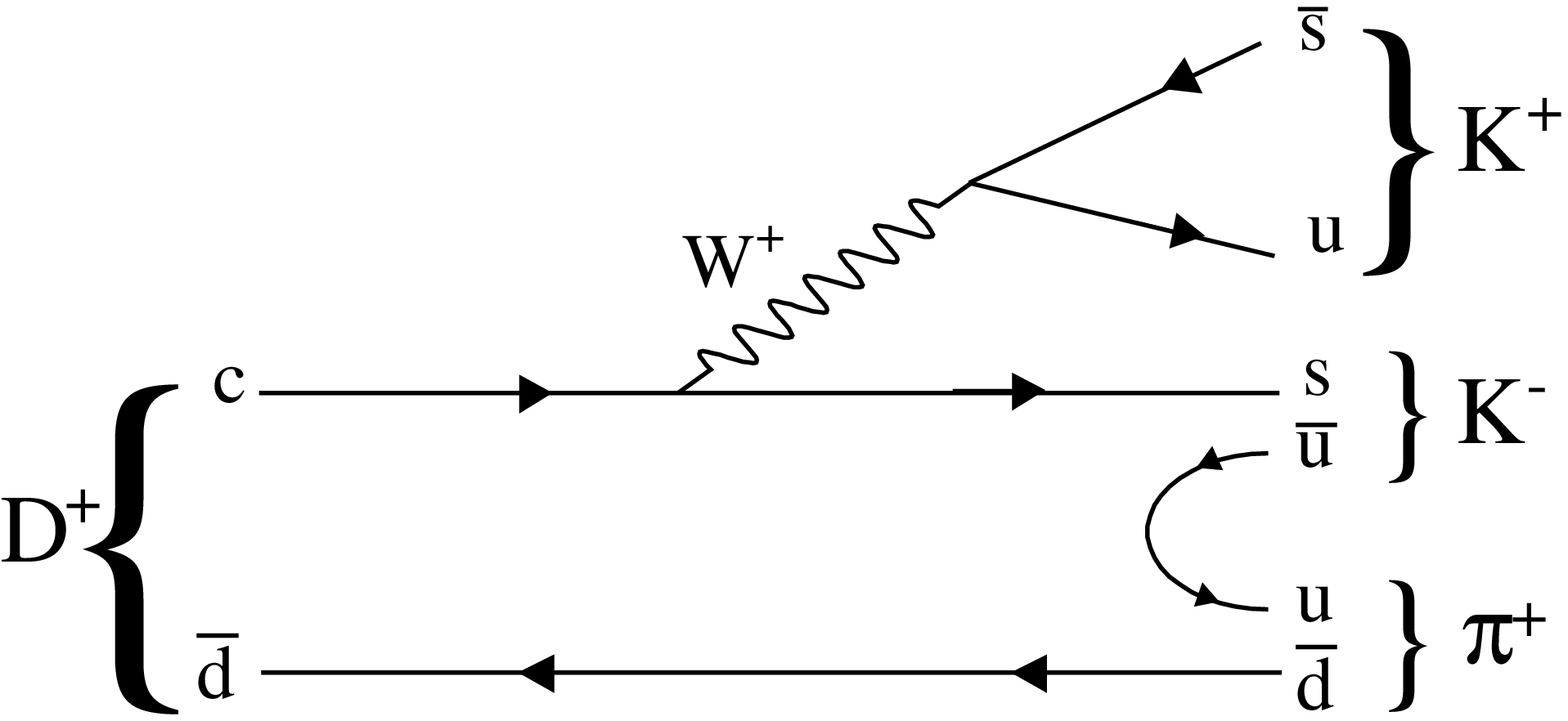}}
  \hfill
  \subfigure[]{\includegraphics[width=.49\columnwidth]{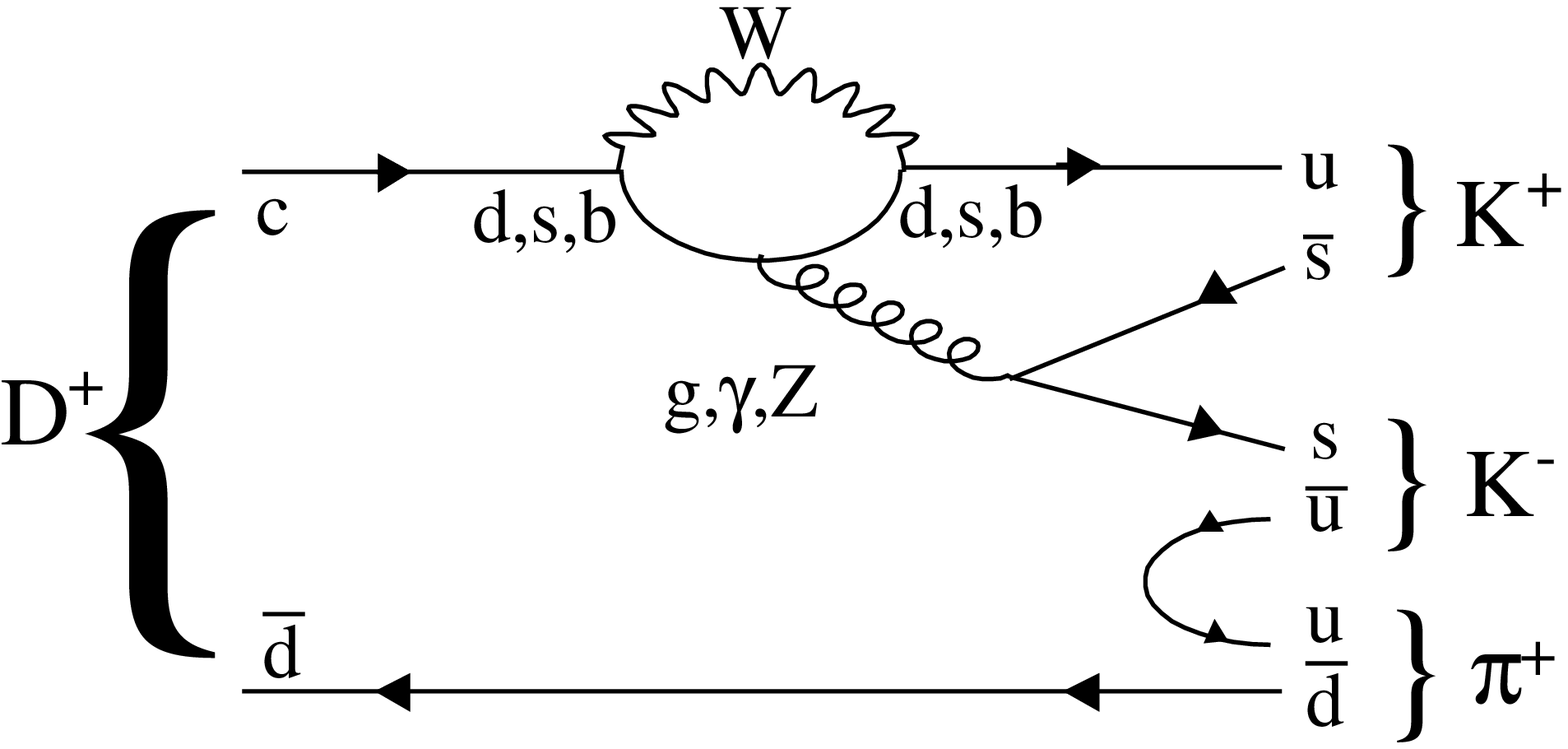}}
    \caption{\label{fig:Feynman}
      Examples of \Dptokkpi decays: (a) a tree diagram, and (b) a penguin process.
    }
\end{figure}

Singly Cabibbo-suppressed (SCS) charged $D$-meson decays are predicted 
in the standard model (SM) to exhibit \CP-violating charge asymmetries of the 
order of $10^{-3}$~\cite{Buccella:1993}.
Direct \CP violation in SCS decays is possible in 
the interference between tree-level, Fig.~\ref{fig:Feynman}(a), and penguin, 
Fig.~\ref{fig:Feynman}(b), decay processes. Doubly Cabibbo-suppressed (DCS) and 
Cabibbo-favored (CF) decays are expected to be \CP invariant in the SM because they 
are dominated by a single weak amplitude. Measurements of \CP asymmetries in SCS processes greater 
than $\cal O$$(10^{-3})$ would be strong evidence of physics beyond the 
standard model~\cite{Bianco}. 

We define the \CP asymmetry by 
\begin{equation}
A_{CP} = \frac{\left|\mathcal{A}\right|^{2}-\left|\overline{\mathcal{A}}\right|^{2}}{\left|\mathcal{A}\right|^{2}+\left|\overline{\mathcal{A}}\right|^{2}},
\label{eq:one}
\end{equation}

\noindent where $\mathcal{A}$ is the total decay amplitude for \Dp decays and $\overline{\mathcal{A}}$ 
is the amplitude for the charge-conjugate decays. $A_{CP}$ is different from 
zero only if there are at least two different decay amplitudes where there has to 
be a relative weak phase and an induced phase shift between the amplitudes due to final-state interaction. 
Eq.~(\ref{eq:one}) can be expressed 
as a function of the branching fractions with the $CPT$ invariance, 
$\Gamma(D^+)=\Gamma(D^-)$. Assuming that CF decays are invariant 
under the \CP transformation, we use them as normalization factors in the asymmetry,
\begin{equation}
A_{CP} = 
 \frac{
   \frac{\mathcal{B}(D^+\rightarrow K^+K^-\pi^+)}
        {\mathcal{B}(D_s^+\rightarrow K^+K^-\pi^+)}
  -\frac{\mathcal{B}(D^-\rightarrow K^+K^-\pi^-)}
        {\mathcal{B}(D_s^-\rightarrow K^+K^-\pi^-)}
 }
 { \frac{\mathcal{B}(D^+\rightarrow K^+K^-\pi^+)}
        {\mathcal{B}(D_s^+\rightarrow K^+K^-\pi^+)}
  +\frac{\mathcal{B}(D^-\rightarrow K^+K^-\pi^-)}
        {\mathcal{B}(D_s^-\rightarrow K^+K^-\pi^-)}
 }.
\label{eqacp}
\end{equation}

\noindent This procedure reduces systematic errors since most of the particle identification (PID) and tracking errors 
cancel out. We also measure the \CP asymmetry in the resonant decays \Dptophipi and \Dptokstark.

Finally, we present a preliminary measurement of the branching ratio $\frac{\Gamma (\Dptokkpi )}{\Gamma(\Dptokpipi)}$.

\section{THE \babar\ DETECTOR AND DATASET}
\label{sec:babar}
This analysis is performed with a data sample recorded on and below the $\Upsilon (4S)$ 
resonance with the \babar\ detector at the \pep2 
asymmetric-energy $e^+e^-$ storage ring. The \babar\ detector is described
in detail in Ref.~\cite{Babar}. The silicon vertex tracker (SVT) and the 
40-layer cylindrical drift chamber (DCH) within a 1.5-T solenoid measure the momenta and energy deposition of 
charged particles. A ring-imaging Cherenkov detector (DIRC) is used for charged-particle identification. Photons are detected and electrons identified 
with a CsI(Tl) electromagnetic calorimeter (EMC). Muons are identified 
in the instrumented flux return (IFR), composed of resistive-plate chambers 
and layers of iron, which return the magnetic flux of the solenoid.

We split our 89.7~\invfb sample into a 9.8~\invfb randomly selected subsample used 
to optimize the selection criteria and the remainder, 79.9~\invfb sample used for the analysis. 
The subsample is used to finalize the analysis procedure including the study of systematic errors. 
This procedure limits selection bias. Furthermore, 
the same selection criteria are applied to the CF and SCS modes whenever possible to
reduce systematic errors. We use 145~\invfb equivalent of (generic) $c\overline{c}$ Monte Carlo (MC)~\cite{G4}  
events to determine efficiencies.

\section{ANALYSIS METHOD}
\label{sec:Analysis}
We reconstruct \Dp and \Ds~\cite{CCsign} mesons by selecting events containing at least three charged tracks. 
Tracks are required to have  at least 12 measured DCH 
coordinates, a minimum transverse momentum of 0.1~\gevc, and to originate within 
1.5~cm in $xy$ (transverse to the beam) and $\pm 10$~cm along the $z$-axis (along the $e^-$ beam) of the nominal interaction point. Kaons are 
identified with a likelihood ratio constructed with 
$\mathrm{d}\hspace{-0.1em}E/\mathrm{d}x$ likelihood functions from the SVT and DCH, and 
a DIRC likelihood function constructed with the Cherenkov angle and number of photons. 
Pions are identified as tracks that fail a loose kaon identification criteria. Three charged 
tracks are fitted constraining their paths to a common vertex, and accepted if the fit probability 
$P(\chi^2)>1\%$. We reject \Dp and \Ds mesons from $B$ decays by requiring that the $D$ 
momentum in the center-of-mass (CM) frame be above 2.4 \gevc.

\begin{figure}
  \vskip -0.15in
  \centering
  \includegraphics[width=.385\columnwidth]{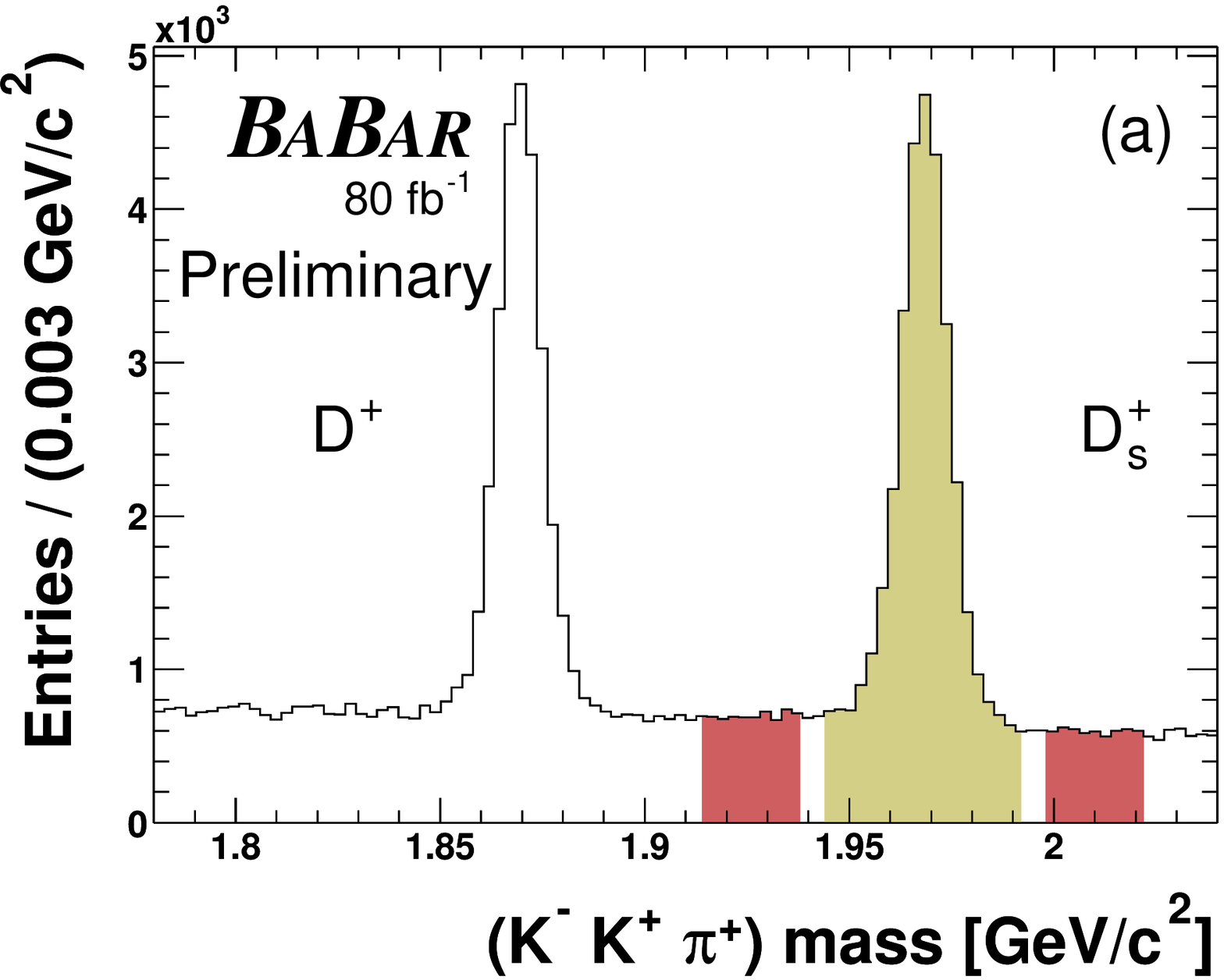} \hspace{1cm}
  \includegraphics[width=.385\columnwidth]{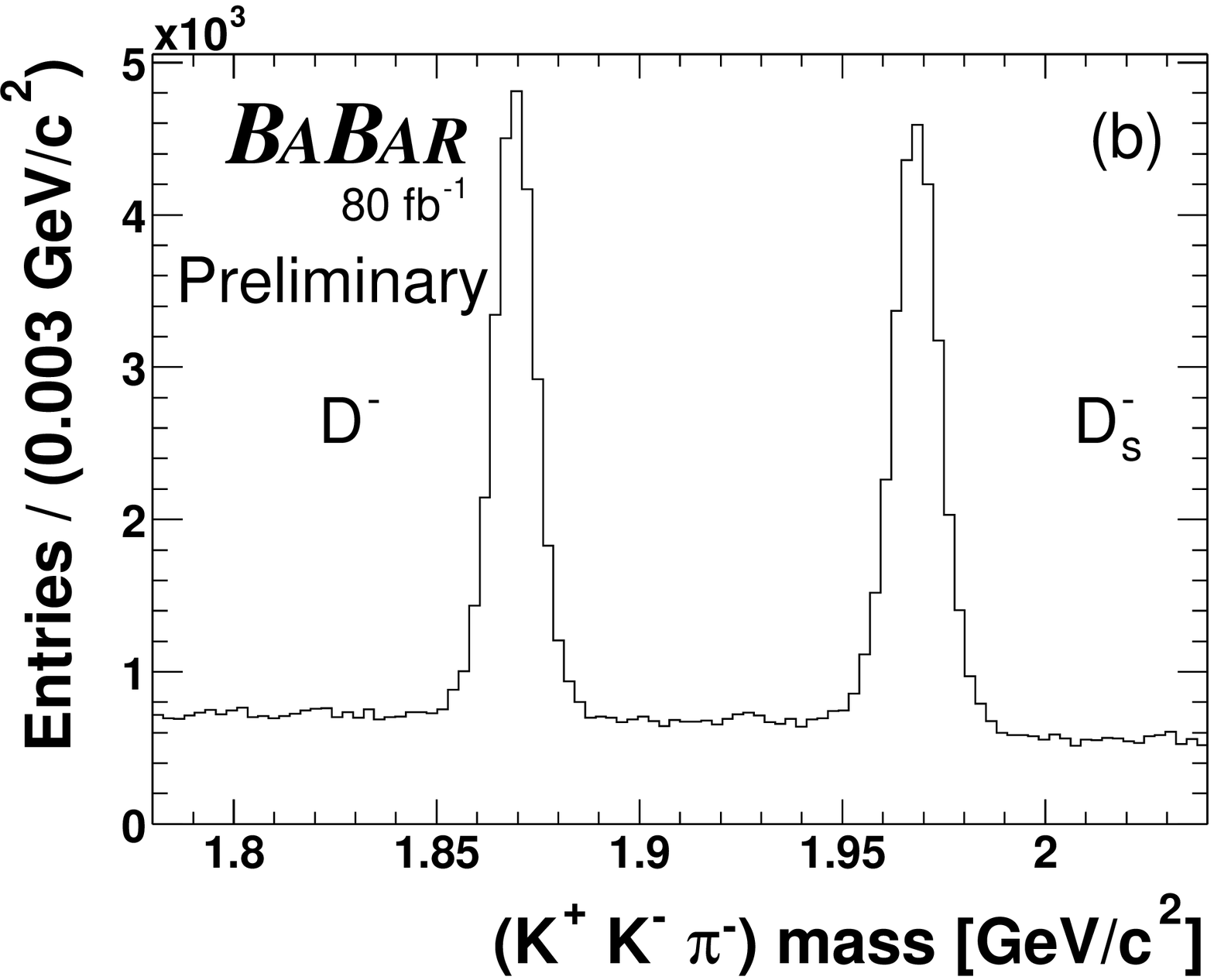}
  \includegraphics[width=.385\columnwidth]{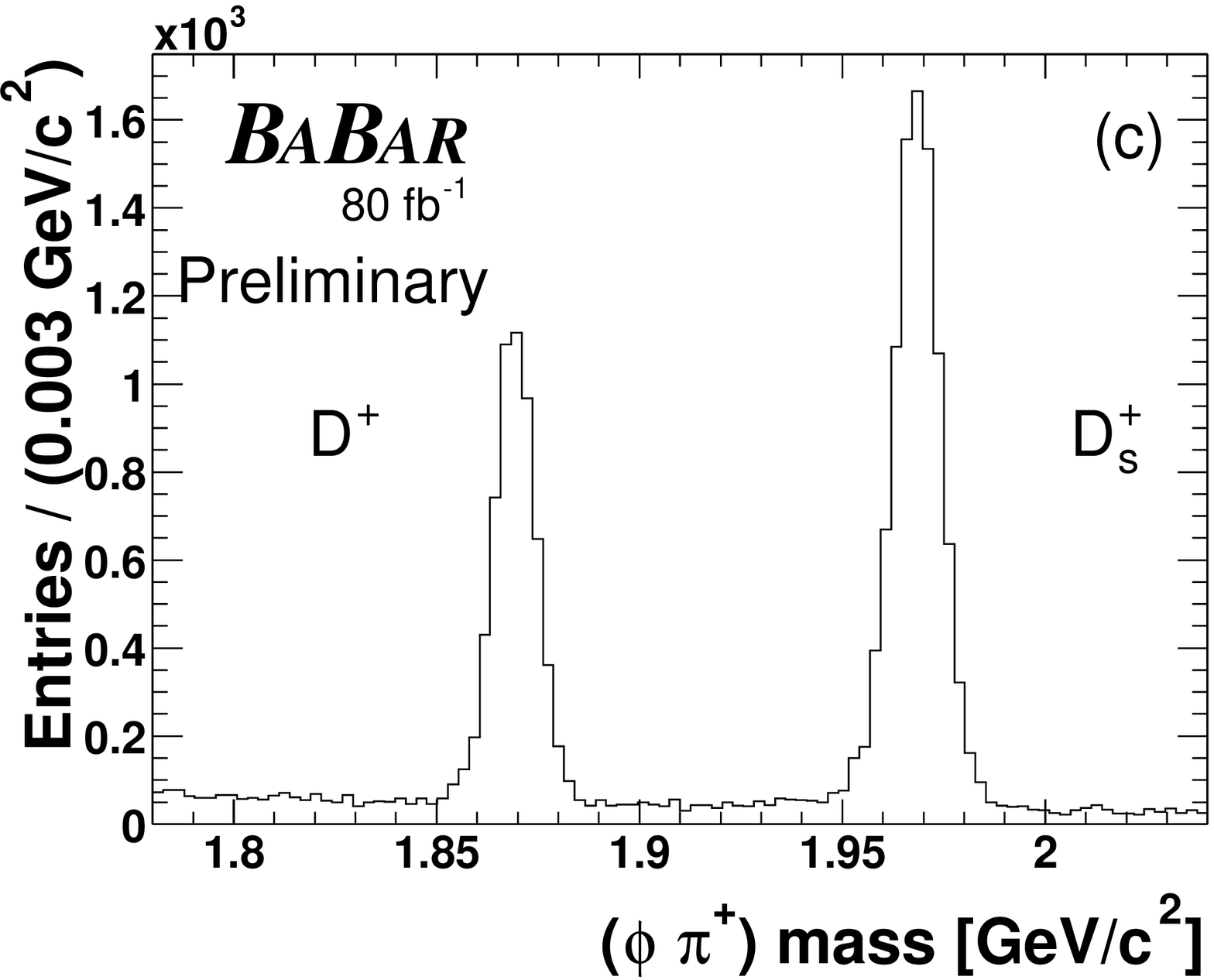} \hspace{1cm}
  \includegraphics[width=.385\columnwidth]{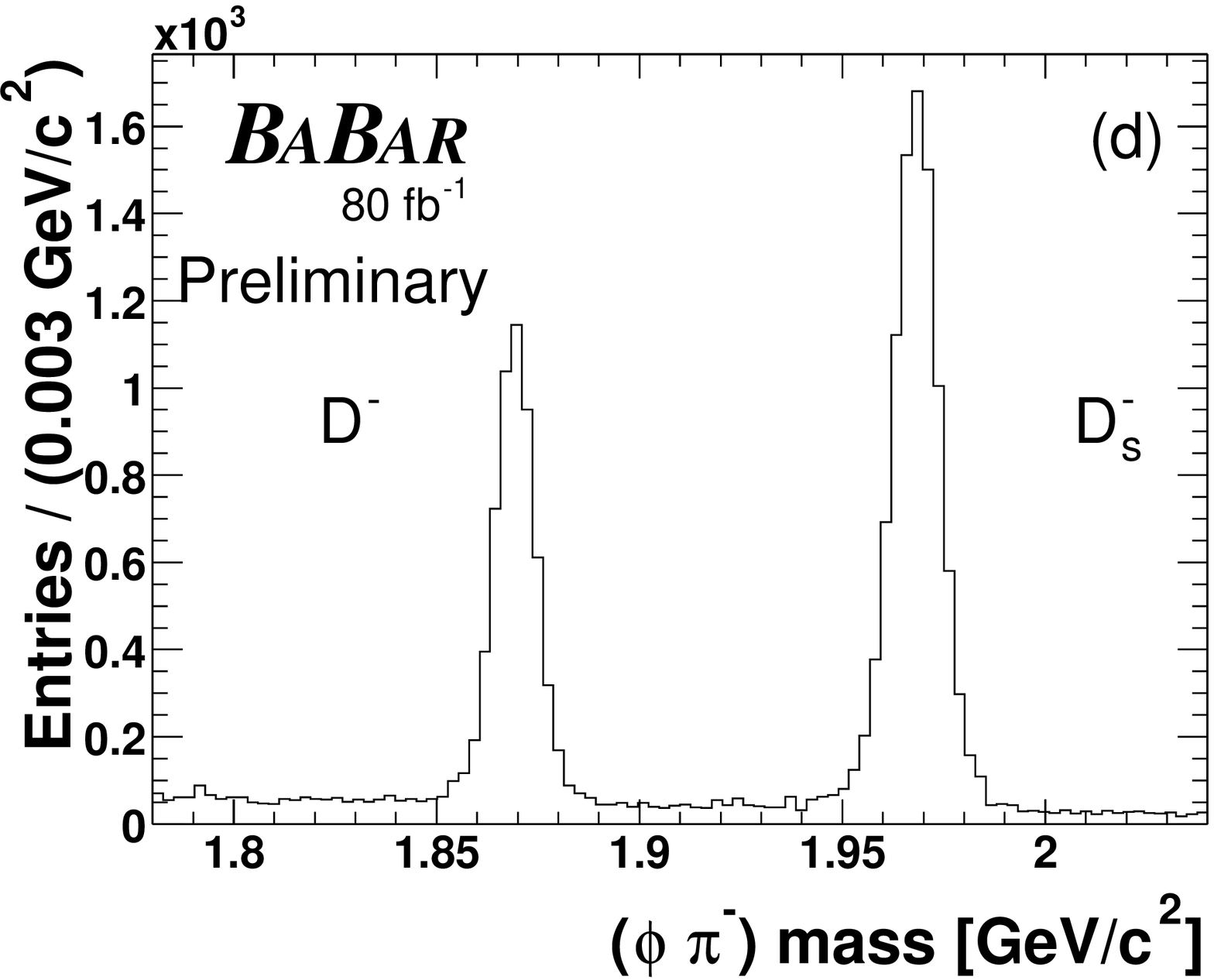}
  \includegraphics[width=.385\columnwidth]{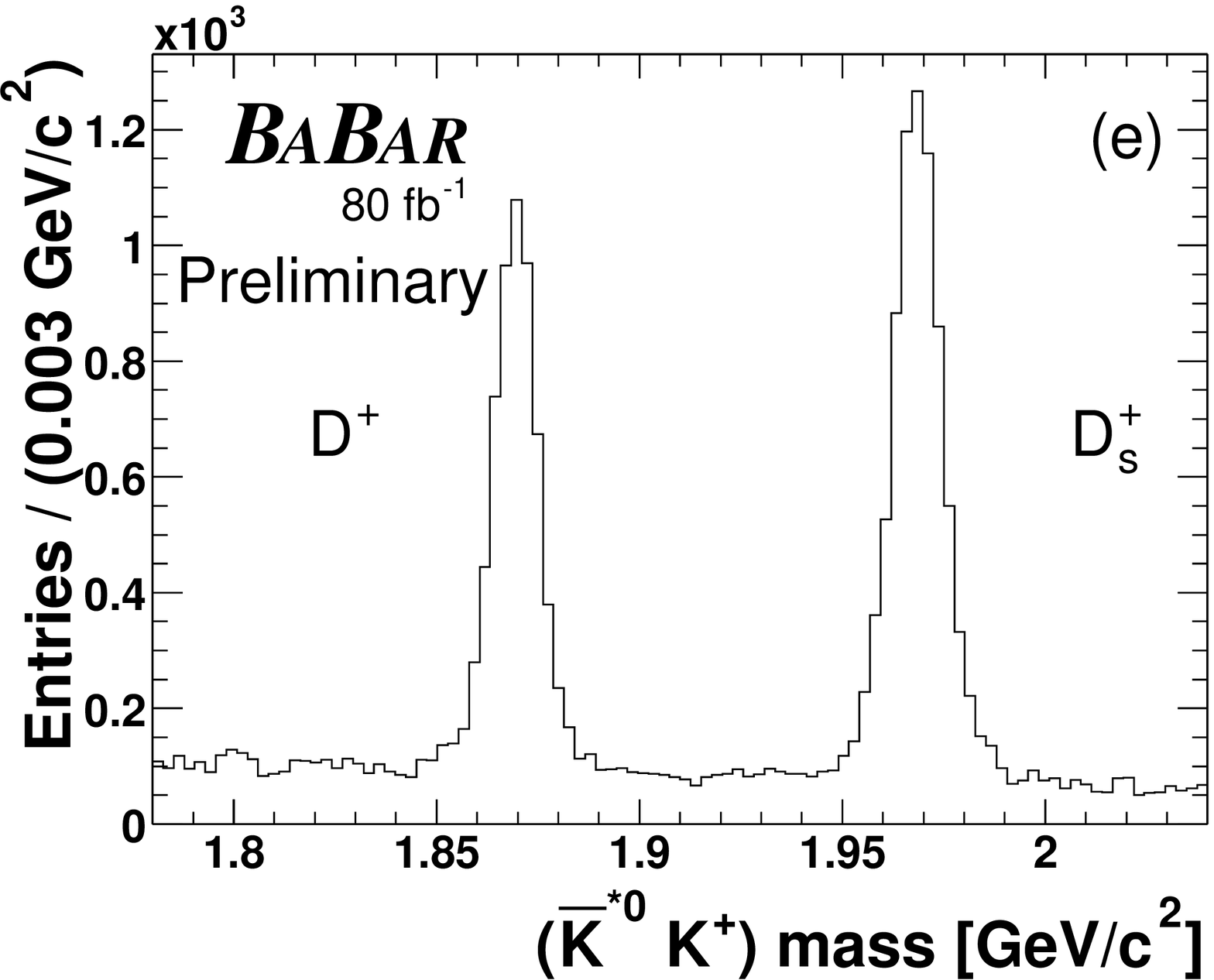} \hspace{1cm}
  \includegraphics[width=.385\columnwidth]{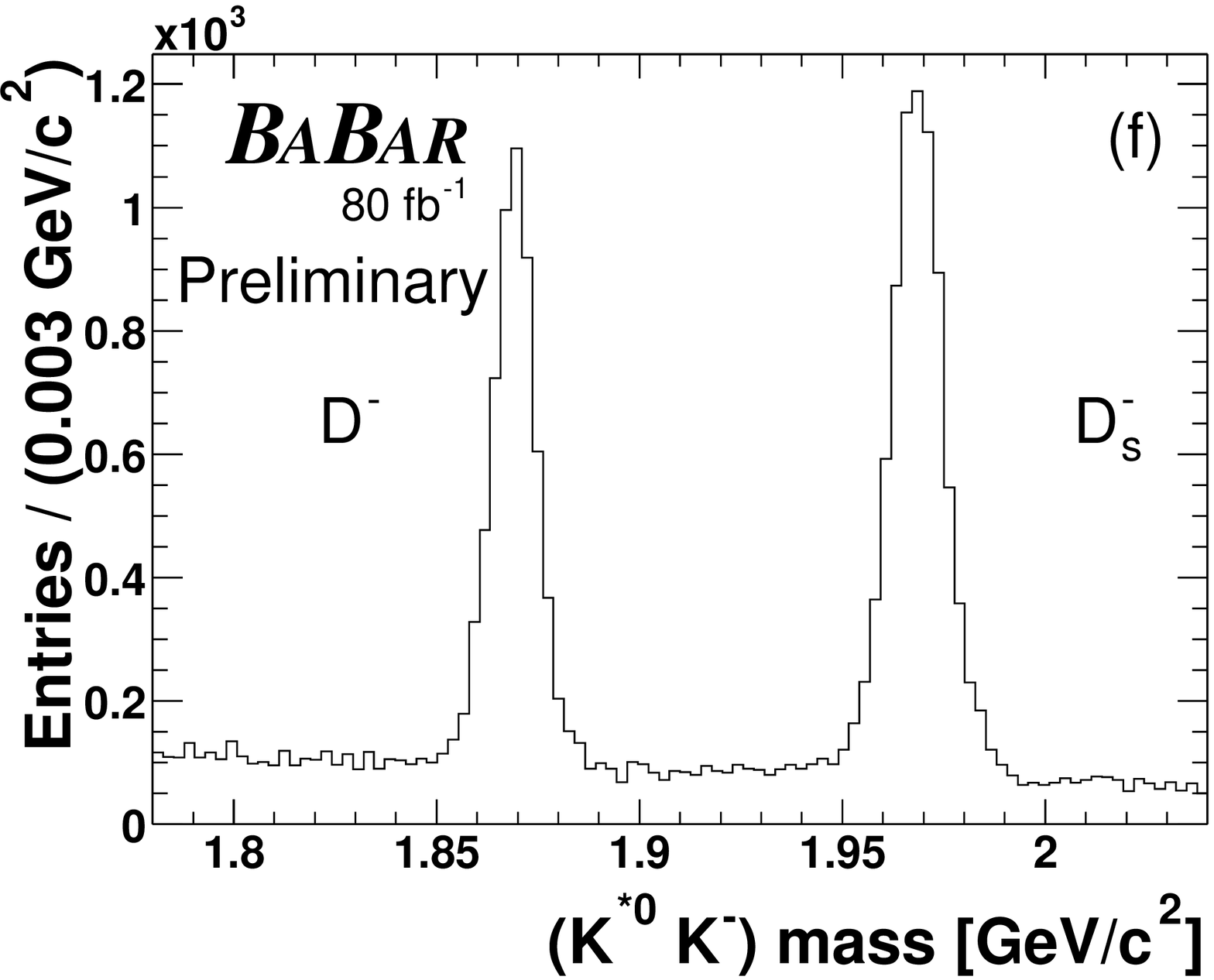}
  \includegraphics[width=.385\columnwidth]{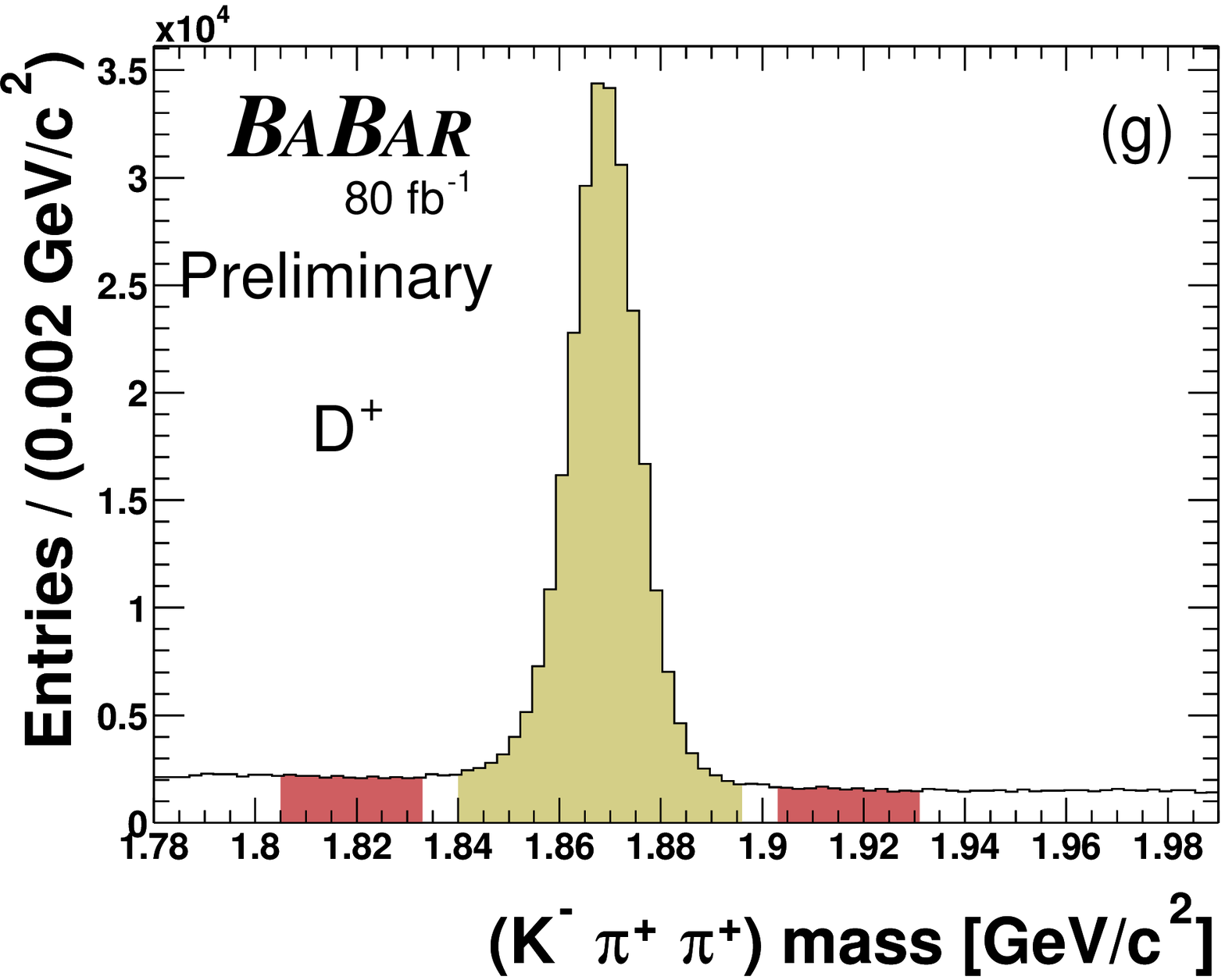} \hspace{1cm}
  \includegraphics[width=.385\columnwidth]{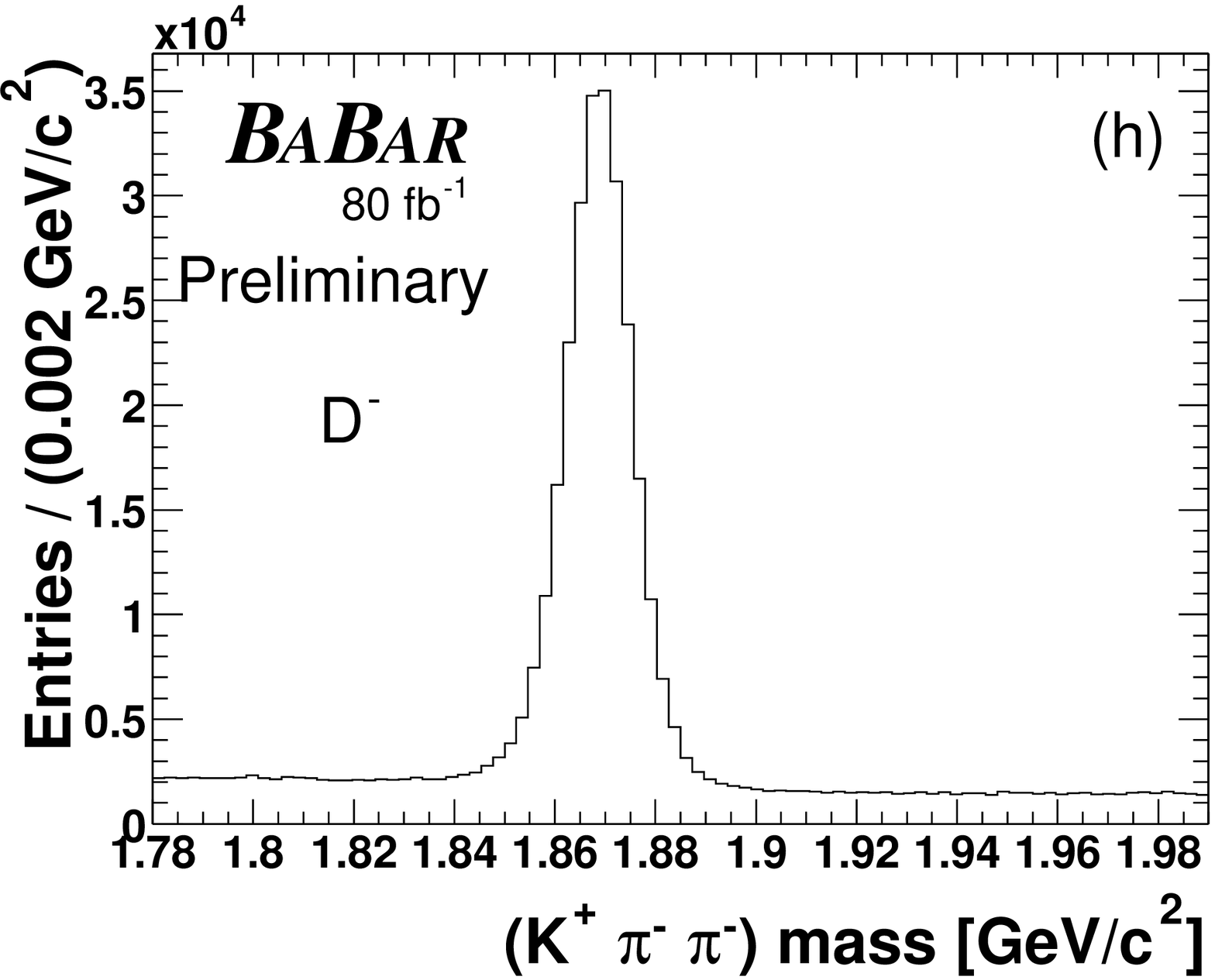}
  \caption{\label{fig:Yields}
    Mass distributions for positively charged (left) and negatively charge (right) $D_{(s)}$ mesons for events 
satisfying the requirement on $r \geq 4.3$. 
Figures (a) and (b) are for all $KK\pi$ candidates, while (c) and (d) are for $\phi \pi$ candidates, and (e) and (f) 
for $\overline{K}^{*0} K$ candidates. Figures (g) and (h) are for $K\pi \pi$ candidates. Signal (yellow or light shaded) 
and sidebands (red or darker shaded) regions are shown for \Ds and \Dp decays in (a) and (g), respectively.
  }
\end{figure}

In order to reduce the remaining combinatorial background we construct a likelihood ratio ($r$) from the probability density functions 
(PDFs) of the following discriminating variables for the \Dp and \Ds 
mesons: CM momentum ($p_{\mathrm{CM}}$) and vertex probability ($\chi^2$-based) 
with beam spot constraint ($P_{\mathrm{BS}}(\chi^2)$). The signal PDFs are obtained with 
a background-subtraction technique from the data subsample. For \Dp decays, the signal band is defined as 
$m_{D^+}\in [1.854,1.882]$ \gevcc and the sideband mass regions as $m_{D^+}\in\{[1.819,1.833]\cup [1.903,1.917]\}$ \gevcc [see Fig.~\ref{fig:Yields} (g)]. A joint 
likelihood function is constructed for the signal, $\mathcal{L}_{\mathrm{sig}} =\prod_{i} \mathcal{L}^{i}_{\mathrm{sig}}(x_i)$, 
and the background, $\mathcal{L}_{\mathrm{bkg}} =\prod_{i} \mathcal{L}^{i}_{\mathrm{bkg}}(x_i)$, where $i$ runs over the variables used. The ratio of the joint likelihoods 
$r=\mathcal{L}_{\mathrm{sig}} / \mathcal{L}_{\mathrm{bkg}}$ 
is a powerful variable to separate signal and background. About 16\% of the events have 
more than one \Dp meson. For such events the likelihood ratio is calculated for each candidate and the 
candidate with the highest likelihood ratio is selected.

The sensitivity $S/\Delta S$, where $S$ is the signal and $\Delta S$ is its error, 
is optimized as a function of the likelihood ratio. Using the subsample, the optimal 
selection is found to be $r \geq 4.3$. This criterion is applied to both CF and SCS decays.

The resonant final states \Dptophipi and \Dptokstark are selected by
requiring that 
the invariant mass of the resonant decays be within 
$0.01$~\gevcc and $0.05$~\gevcc of the nominal $\phi$ and \Kstarzb masses, respectively. 
In addition, the signal is optimized by a selection on the cosine 
of the helicity angle ($\cos \theta_H$). 
In the \Dptophipi decay mode, the helicity angle is defined as the angle
between the  $K^{-}$  and
the  $\pi^{+}$  in the  $\phi$  rest frame. In the \Dptokstark decay mode, the helicity angle is defined as the
angle between the  $K^{-}$  from the $\overline{K}^{*0}$  and the  $K^{+}$  in the
$\overline{K}^{*0}$ rest frame. Maximum sensitivity is obtained when $| \cos \theta_H | \geq 0.2$ 
and $| \cos \theta_H | \geq 0.3$ for \Dptophipi and \Dptokstark, respectively.

The CF \Dsptokkpi decays are selected similarly. The signal 
and sideband mass regions are chosen to be  $m_{KK\pi}\in[1.944,1.992]$~\gevcc, and 
$m_{KK\pi}\in\{[1.914,1.938]\cup[1.938,1.998]\}$~\gevcc, respectively [see Fig.~\ref{fig:Yields} (a)]. The only difference from the 
SCS case is that contamination from \Dptokpipi decays is removed. In the 
$KK\pi$ candidates, the kaon with the same charge as the pion is labeled 
as a pion and then the $K\pi\pi$ invariant mass is calculated. We observe a 
\Dp peak indicating that part of the \Ds signal is composed of misidentified \Dp 
candidates. Events in the region $1.855\leq m_{K\pi\pi} \leq 1.883$~\gevcc are removed 
from the \Ds sample.

Contamination from $D^{*+}\rightarrow D^0(\rightarrow K^-\pi^+,K^-K^+)\pi^+$
decays is removed with 
a kinematic requirement on the $D^0$ invariant mass, $m_{K^-h^+}\geq 1.84$~\gevcc. 
In the case of the \Dptokpipi decays, both $K\pi$ combinations must satisfy the 
requirement. Partially reconstructed $D^0\rightarrow K^-\pi^+\pi^0$ events are 
also misidentified as \Dp events when the $\pi^0$ is missed and the charged pion 
is misidentified as a kaon. Most of these events are removed by labeling
a kaon track as a pion and applying 
a restriction on the mass difference $0.139 \leq (m_{K^-\pi^+\pi^+} -
m_{K^-\pi^+}) \leq 0.150$~\gevcc.

The optimized selection criteria are applied to the final sample to obtain 
the signal yields. Figure~\ref{fig:Yields} shows the mass
distributions. The yields, listed in Table~\ref{tbl:yields}, are computed by 
subtracting a scaled background estimate obtained from the sideband mass regions from the 
number of events in the signal region. This technique minimizes sensitivity to background shape 
assumptions.

\begin{table}[htb]
  \caption{\label{tbl:yields}
    Summary of yields in signal and normalization modes.
  }
  \begin{center}
    \begin{tabular}{lcc}
      \hline\hline
      \multicolumn{1}{c}
      {Parent Charge} &     $+$      &         $-$      \\
      \hline

      \Dptokkpi   & $ 21632 \pm 228$ & $ 20940 \pm 226$ \\
      \Dptophipi  & $  5452 \pm  87$ & $  5327 \pm  86$ \\
      \Dptokstark & $  5247 \pm  96$ & $  5113 \pm  96$ \\
      \Dsptokkpi  & $ 23066 \pm 217$ & $ 22928 \pm 214$ \\
      \Dptokpipi  & $236254 \pm 570$ & $237616 \pm 571$ \\

      \hline\hline
    \end{tabular}
  \end{center}
\end{table}

The efficiencies needed for the $A_{CP}$ calculation are obtained from a sample of 
MC (generic) $c\overline{c}$ events. The selection criteria
optimized for the subsample are applied to the MC sample. Efficiencies 
are then calculated as ratios of the numbers of selected signal MC events to numbers 
of generated events. The decay efficiencies are shown 
in Table~\ref{tbl:Efficiencies}. 

We use Eq.~(\ref{eqacp}) (recognizing that branching fractions are
proportional to yields divided by efficiencies) to obtain
$A_{CP}$. As cross-checks, we calculate the \CP asymmetries with two other methods 
(which would have larger systematic errors than the primary method): (i) using the CF \Dptokpipi 
branching fraction as normalization, $A^{(1)}_{CP}$, and (ii) without any normalization factor, 
$A^{(2)}_{CP}$, and the results are shown in Table~\ref{tbl:Acp}.
A study of the \CP asymmetry in bins of the \Dptokkpi Dalitz plot
indicates that the asymmetry is consistent with being constant and zero.

The relative branching ratio 
$\frac{\Gamma (\Dptokkpi )}{\Gamma(\Dptokpipi)}$ 
has been calculated as follows. The CF and SCS 
Dalitz plots are first binned to have equally populated bins. 
Then, the signal and normalization yields and efficiencies 
are calculated bin by bin. The efficiency-corrected yields are then summed 
and divided to obtain the ratio. We obtain a branching ratio of
$(10.7\pm0.1(\textrm{stat.}))\times 10^{-2}$ with the final
sample.

\begin{table}[htb]
  \caption{\label{tbl:Efficiencies}
    Summary of the efficiencies for positively ($\varepsilon^+$) and negatively ($\varepsilon^-$) 
charged $D_{(s)}$-meson decays. Efficiencies are in percent [\%].
  }
  \begin{center}
    \begin{tabular}{lr@{$\pm$}lr@{$\pm$}l}
      \hline\hline
      \multicolumn{1}{c}{Decay} & \multicolumn{2}{c}{$\varepsilon^+$} & \multicolumn{2}{c}{$\varepsilon^-$} \\
      \hline

      \Dptokkpi   & 8.20&0.04 & 8.26&0.04 \\
      \Dptophipi  & 7.67&0.07 & 7.63&0.07 \\
      \Dptokstark & 5.88&0.07 & 5.90&0.07 \\
      \Dptokpipi  & 9.90&0.02 & 10.17&0.02 \\
      \Dsptokkpi  & 3.77&0.02 & 3.79&0.02 \\

      \hline\hline
    \end{tabular}
  \end{center}
\end{table}

\begin{table}[htb]
  \caption{\label{tbl:Acp}
    Summary of \CP asymmetries measured in three different ways.
  }
  \begin{center}
    \begin{tabular}{lccc}
      \hline\hline
      \multicolumn{1}{c}{} & $A_{CP}~[10^{-2}]$ & $A^{(1)}_{CP}~[10^{-2}]$ & $A^{(2)}_{CP}~[10^{-2}]$ \\
      \hline

      $(K^-K^+\pi^{\pm})$  & $+1.36\pm1.01$ & $+0.58\pm0.86$ & $+2.07\pm0.84$ \\
      $(\phi \pi^{\pm})$   & $+0.24\pm1.45$ & $-0.54\pm1.35$ & $+0.94\pm1.33$ \\
      $(\Kstarzb K^{\pm})$ & $+0.88\pm1.67$ & $+0.10\pm1.58$ & $+1.58\pm1.57$ \\

      \hline\hline
    \end{tabular}
  \end{center}
\end{table}

\section{SYSTEMATIC ERRORS AND CROSS-CHECKS}
\label{sec:Systematics}

We estimate the systematic error on the \CP asymmetries in three
different ways. The first approach combines estimates of the contributions from various
identified sources listed in Table~\ref{tbl:Syst_Acp}.  The second and third estimates
come from partly redundant direct studies of asymmetries in the
normalization and control samples.

The first row of Table~\ref{tbl:Syst_Acp} gives the error ($0.06$\%) assigned to $A_{CP}$ 
due to small differences in momentum spectra of $\pi$, $K$ from \Dp and \Ds decays. 
This error is estimated as three times the maximum difference in $\pi$, $K$ MC-efficiency 
asymmetries in \Dp and \Ds decays. We evaluate an error for the
background subtraction by changing the widths of the sideband mass regions. The
error is taken to be the difference in the central values of
$A_{CP}$. The errors in the likelihood-ratio technique are estimated
with two variants: (i) tightening the likelihood ratio to produce a
$10$\% change in the decay yields, and (ii) using another likelihood
ratio ($r_1$) which incorporates a third variable, the distance in the $xy$-plane from the 
interaction point to the \Dp vertex ($d_{xy}$). Case (i) is obtained at $r \geq 6.0$ and the maximum
sensitivity in case (ii) is at $r_1 \geq 8.8$. Table~\ref{tbl:Syst_Acp}
summarizes these systematic errors for the \CP asymmetries. The total errors are 0.8\%, 0.7\%, 
and 0.7\% on the \Dptokkpi, \Dptophipi, and \Dptokstark asymmetries, respectively.

Our second estimate is the larger of the differences between $A_{CP}$
and the other two measurements $A_{CP}^{(1)}$ and $A_{CP}^{(2)}$. The error  
is $0.8\%$ on both the inclusive \Dptokkpi and for the resonant asymmetries.

Our third and final estimate, which is applicable only to the inclusive
three-body final states, is based on \CP asymmetries for the CF decays
\Dptokpipi and \Dsptokkpi (since these are expected to be zero within the SM). 
The high-statistics \Dptokpipi control mode is used to search for the scale of systematic
effects intrinsic to the detector. The \Dsptokkpi mode, which is also our 
primary normalization mode, is largely
insensitive to these effects since it has the same final state as our
signal decay. In \Dsptokkpi decays, both the $D_s^+$ and the
$D_s^-$ decay to two oppositely charged kaons; only pion charge differs
in particle and antiparticle decays. In \Dptokpipi decays
however, all three particles have opposite charges in particle and
antiparticle decays. For these control samples, we obtain asymmetries 
of $(+1.1\pm0.2)\times 10^{-2}$ and $(+0.6\pm0.8)\times 10^{-2}$ (statistical errors only) 
for \Dptokpipi and \Dsptokkpi, respectively.

We find that these observed control-sample asymmetries are almost entirely in
the efficiencies derived from our 
simulation of the detector, rather than the signal yields (see Tables~\ref{tbl:yields} and~\ref{tbl:Efficiencies}). 
With corrections based on estimates of low-energy nuclear interaction effects which are 
not accounted for in the simulation we find the smaller asymmetries 
$+0.8\times 10^{-2}$ and $+0.4\times 10^{-2}$ for \Dptokpipi and \Dsptokkpi
, respectively. These results we interpret to mean
that our present simulation of particle interactions in the material of the detector is incomplete.

Even though our definition of
$A_{CP}$ [see Eq.~(\ref{eqacp})] invokes a normalization chosen to eliminate effects
that may be important here, we choose here the conservative estimate
that the systematic error is measured by the magnitude of the departure
of these CF asymmetries from the expected null values.

We chose as our systematic error in the \CP
asymmetry the largest of all applicable estimates, 1.1\% on \Dptokkpi and 0.8\% on the resonant decays.

The \CP asymmetry has been cross-checked as a function of the \Dp lab momentum as well as 
by the year of data production. The $\chi^2$-based
probability of the asymmetry to be constant is $32$\% and $63$\% for
momentum and time-period dependences, respectively.

A summary of the systematic errors for the branching ratio 
$\frac{\Gamma (\Dptokkpi )}{\Gamma(\Dptokpipi)}$ 
is given in Table~\ref{tbl:Syst_BR}. The fractional error due to PID and
tracking has been estimated as 2.1\% of the branching ratio, computed as 
the sum in quadrature of 1.1\%\ for PID and 1.8\%\ for tracking~\cite{Ups4S}.

\begin{table}[htb]
  \caption{
    \label{tbl:Syst_Acp}
    Summary of systematic errors for the \CP asymmetries.
  }
  \begin{center}
    \begin{tabular}{lccc}
      \hline\hline
      \multicolumn{1}{c}{Source} & $(K^-K^+\pi^{\pm})$ & $(\phi \pi^{\pm})$ & $(\Kstarzb K^{\pm})$ \\
      \multicolumn{1}{c}{} & $A_{CP}~[10^{-2}]$ & $A_{CP}~[10^{-2}]$ & $A_{CP}~[10^{-2}]$ \\
      \hline

      MC simulation        & $0.06$   & $0.06$   & $0.06$ \\
      Background estimate  & $0.63$   & $0.32$   & $0.49$ \\
      Using $r \geq 6.0$   & $0.22$   & $0.15$   & $0.01$ \\
      Using $r_1 \geq 8.8$ & $0.46$   & $0.54$   & $0.54$ \\
      \hline
      Total                & $0.81$   & $0.65$   & $0.73$ \\
      \hline\hline

    \end{tabular}
    \end{center}
\end{table}

\section{RESULTS and SUMMARY}
\label{sec:ResSum}

In summary, we have searched for a \CP asymmetry in 
\Dptokkpi, \Dptophipi, and \Dptokstark decays and measured the
branching ratio of \Dptokkpi decays all with a data sample of $79.9$ \invfb collected by the \babar\ experiment.

Preliminary measurements of the \CP asymmetries are 
$A_{CP}(K^{+}K^{-}\pipm) = (1.4\pm1.0(\textrm{stat.})\pm1.1(\textrm{syst.}))\times 10^{-2}$,
$A_{CP}(\mphi\pipm) = (0.2\pm1.5(\textrm{stat.})\pm0.8(\textrm{syst.}))\times 10^{-2}$, and
$A_{CP}(K^{\pm}\Kstarzb) = (0.9\pm1.7(\textrm{stat.})\pm0.8(\textrm{syst.}))\times 10^{-2}$.
These results are in agreement with previous published
results~\cite{refcp}, with our results in the resonant modes having
significantly smaller errors. 

Further, we obtain a preliminary branching ratio for \Dptokkpi decays
relative to \Dptokpipi decays of
$(10.7\pm0.1(\textrm{stat.})\pm0.2(\textrm{syst.}))\times 10^{-2}$. This result is a
significant improvement over previous measurements~\cite{refbr}. 

%
%

\begin{table}[htb]
  \caption{\label{tbl:Syst_BR}
    Systematic errors for the branching ratio.
  }
  \begin{center}
    \begin{tabular}{lcc}
      \hline\hline
      \multicolumn{1}{c}{Source} & Error~$[10^{-2}]$ \\
      \hline

      PID + tracking         & $0.22$ \\
      Background estimate    & $0.05$ \\
      Using $r \geq 6.0$     & $0.00$ \\
      Using $r_1 \geq 8.8$   & $0.02$ \\
      \hline
      Total                  & $0.23$ \\

      \hline\hline
    \end{tabular}
  \end{center}
\end{table}

\section{ACKNOWLEDGMENTS}
\label{sec:Acknowledgments}

\input pubboard/acknowledgements

\end{document}

%% file: pubboard/authors_sum2004.tex
\begin{center}
\small

The \babar\ Collaboration,
\bigskip

%
B.~Aubert,
R.~Barate,
D.~Boutigny,
F.~Couderc,
J.-M.~Gaillard,
A.~Hicheur,
Y.~Karyotakis,
J.~P.~Lees,
V.~Tisserand,
A.~Zghiche
\inst{Laboratoire de Physique des Particules, F-74941 Annecy-le-Vieux, France }
% ...

\end{center}\newpage

%% file: pubboard/acknowledgements.tex
We are grateful for the 
extraordinary contributions of our \pep2\ colleagues in
achieving the excellent luminosity and machine conditions
that have made this work possible.
The success of this project also relies critically on the 
expertise and dedication of the computing organizations that 
support \babar.
The collaborating institutions wish to thank 
SLAC for its support and the kind hospitality extended to them. 
This work is supported by the
US Department of Energy
and National Science Foundation, the
Natural Sciences and Engineering Research Council (Canada),
Institute of High Energy Physics (China), the
Commissariat \`a l'Energie Atomique and
Institut National de Physique Nucl\'eaire et de Physique des Particules
(France), the
Bundesministerium f\"ur Bildung und Forschung and
Deutsche Forschungsgemeinschaft
(Germany), the
Istituto Nazionale di Fisica Nucleare (Italy),
the Foundation for Fundamental Research on Matter (The Netherlands),
the Research Council of Norway, the
Ministry of Science and Technology of the Russian Federation, and the
Particle Physics and Astronomy Research Council (United Kingdom). 
Individuals have received support from 
CONACyT (Mexico),
the A. P. Sloan Foundation, 
the Research Corporation,
and the Alexander von Humboldt Foundation.

%% file: conf_sum2004.bbl
\begin{thebibliography}{99}
\bibitem{CCsign}
  Charge conjugation is assumed throughout the selection unless 
otherwise indicated.
\end{thebibliography}
